\documentclass[%
notitlepage, 
lengthcheck, 
 groupedaddress,
 amsmath,amssymb,
 prl,
]{revtex4-1}
\usepackage{siunitx}
\usepackage{braket}
\usepackage{graphicx}
\usepackage{xspace}

\newcommand*{\ie}{i.e.\@\xspace}

\begin{document}
\title{Single-Mode Optical Waveguides on Native High-Refractive-Index Substrates}

\author{Richard R. Grote, Lee C. Bassett$^*$}

\affiliation{Quantum Engineering Laboratory, Department of Electrical and Systems Engineering, University of Pennsylvania, 200 S. 33rd Street, Philadelphia, PA 19104, USA}

\affiliation{$^*$Corresponding author: lbassett@seas.upenn.edu}


\maketitle

{\bf High-refractive-index semiconductor optical waveguides form the basis for modern photonic integrated circuits (PICs) , but the conventional methods of achieving optical confinement require a thick lower-refractive-index support layer that impedes large-scale co-integration with electronics.  To address this challenge, we present a general architecture for single-mode waveguides that confine light in a high-refractive-index material on a native substrate. Our waveguide consists of a high-aspect-ratio fin of the guiding material surrounded by lower-refractive-index dielectrics and is compatible with standard top-down fabrication techniques.  The proposed waveguide geometry removes the need for a buried-oxide-layer in silicon photonics, as well as the InGaAsP layer in InP-based PICs and will allow for photonic integration on emerging material platforms such as diamond and SiC. } 

PICs are rapidly being developed for high-refractive index materials that allow for tight optical confinement, small on-chip bend radii, and strong light-matter interactions.  For example, high-performance PICs in both silicon \cite{Reed_08} and InP \cite{Coldren_JLT_11, Smit_SST_14} platforms are playing an increasingly important role in data applications with the potential to enable exascale computing \cite{Rumley_JLT_15} and on-chip core-to-core optical communication \cite{Kimerling_MRS_14}.  Similarly, wide-band gap semiconductors, such as diamond \cite{Aharonovich_NP_11, Castelletto_NJP_11, Faraon_NP_11, Burek_NComm_14, Thomas_OpEx_14, Mouradian_PRX_15, Hausmann_NP_14, Latawiec_Optica_15} and SiC \cite{Cardenas_OL_15}, have emerged as promising materials for a plethora of new PIC applications.  Among these are non-linear optics \cite{Hausmann_NP_14, Latawiec_Optica_15, Cardenas_OL_15} and integrated quantum information processing \cite{Faraon_NP_11, Burek_NComm_14, Thomas_OpEx_14, Mouradian_PRX_15, Gao_NP_15}, which is enabled by the presence of spin defects with desirable quantum properties \cite{Weber_PNAS_10, Awschalom_Science_13}.  

Common to all of these applications is a need for low-propagation-loss single-mode waveguides that can be fabricated on a high-refractive-index substrate in a scalable fashion.  While a high refractive index is beneficial for optical design, it also requires a buried lower-refractive-index layer and the transfer or growth of thin films of high-index material \cite{Reed_08, Sun_JSSC_15, Coldren_JLT_11, Smit_SST_14, Aharonovich_NP_11, Castelletto_NJP_11, Faraon_NP_11, Hausmann_NP_14, Thomas_OpEx_14, Latawiec_Optica_15, Cardenas_OL_15}, free-standing structures \cite{Orcutt_OpEx_11, Burek_NComm_14, Mouradian_PRX_15}, or pedestals \cite{Lin_OL_13} to minimize optical power leakage from the waveguide into the substrate.  These approaches limit the device robustness, uniformity, and scalability required for the development of dense PICs on wide-bandgap semiconductors. 

Even on mature PIC platforms optical confinement presents significant technological challenges.  In silicon photonics, the buried-oxide-layer thickness required for optical confinement is much larger than the optimum for VLSI electronics, making co-integration difficult \cite{Orcutt_OpEx_11, Kimerling_MRS_14, Sun_JSSC_15}. For InP-based PICs, optical confinement is limited by the low index contrast between InP and InGaAsP \cite{Coldren_JLT_11, Smit_SST_14}.  

\begin{figure}[t!]
\centering
\includegraphics[scale=1]{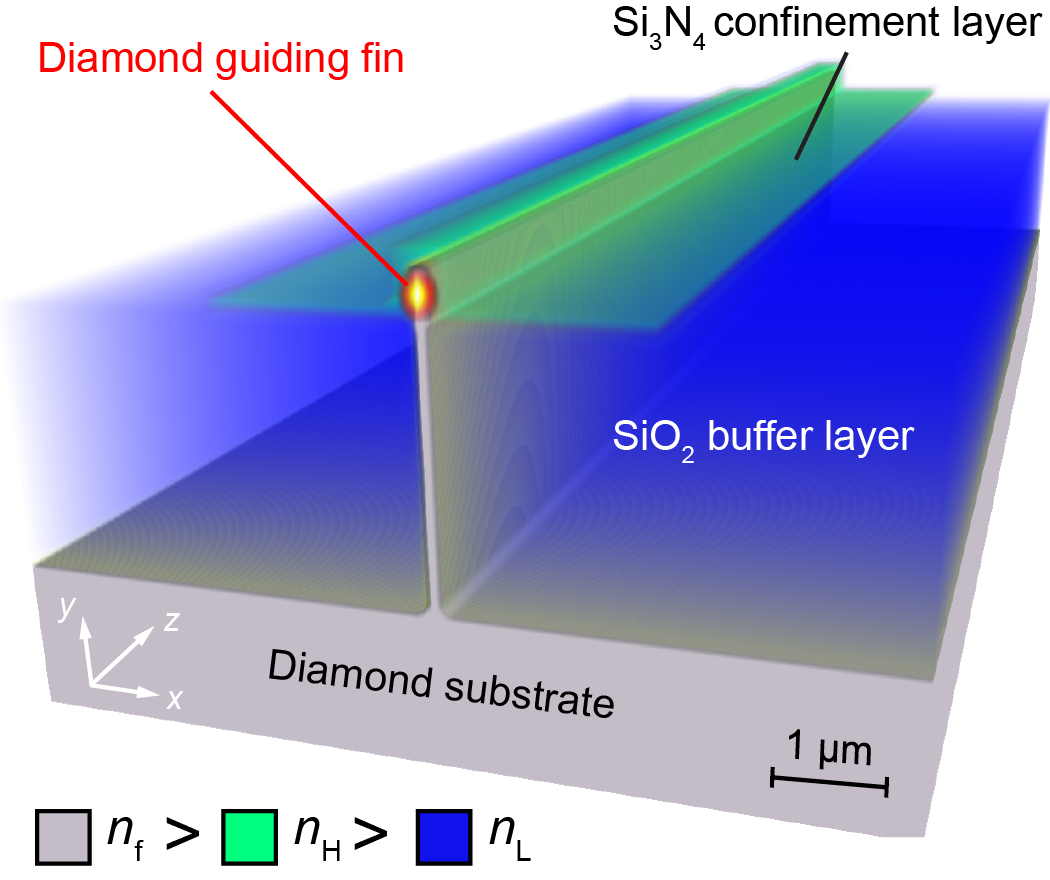}
\caption{\textbf{The fin waveguide.} An example of a fin waveguide on a diamond substrate designed for $\lambda = $~\SI{637}{\nano\meter}.  The geometry supports a single mode when $n_{\text{f}} > n_{\text{H}} > n_{\text{L}}$. The profile of the confined optical mode shown in cross-section has been calculated numerically.}
\label{fig1}
\end{figure}

Here, we propose a new type of waveguide optimized for high-index substrates that utilizes stacked dielectric layers to confine light in the top of a fin of high-index material.  An example of a SiO$_2$/Si$_3$N$_4$ stack on a diamond fin/substrate at a wavelength of $\lambda = $~\SI{637}{\nano\meter} is shown in Fig.~\ref{fig1}.  Although the refractive index of both the buffer and confinement layers ($n_{\text{L}}=n_{\text{SiO}_2} \approx 1.45$ and $n_{\text{H}}=n_{\text{Si}_3\text{N}_4} \approx 2.0$, respectively) are lower than that of the fin and substrate ($n_{\text{f}}=n_{\text{diamond}} \approx 2.4$), the proposed design achieves confinement by engineering the \emph{effective index}, resulting in an optical mode confined within the high-index material (diamond in the case of Fig.~\ref{fig1}). This waveguide mode can propagate without leaking power into the underlying substrate, while the waveguide itself can be fabricated using conventional top-down lithography, etching, dielectric deposition, and planarization techniques.  Our proposed architecture obviates the need for a buried low-index layer, providing a pathway towards large-area, scalable PICs on native substrates.

\begin{figure}[t!]
\centering
\includegraphics[scale=1]{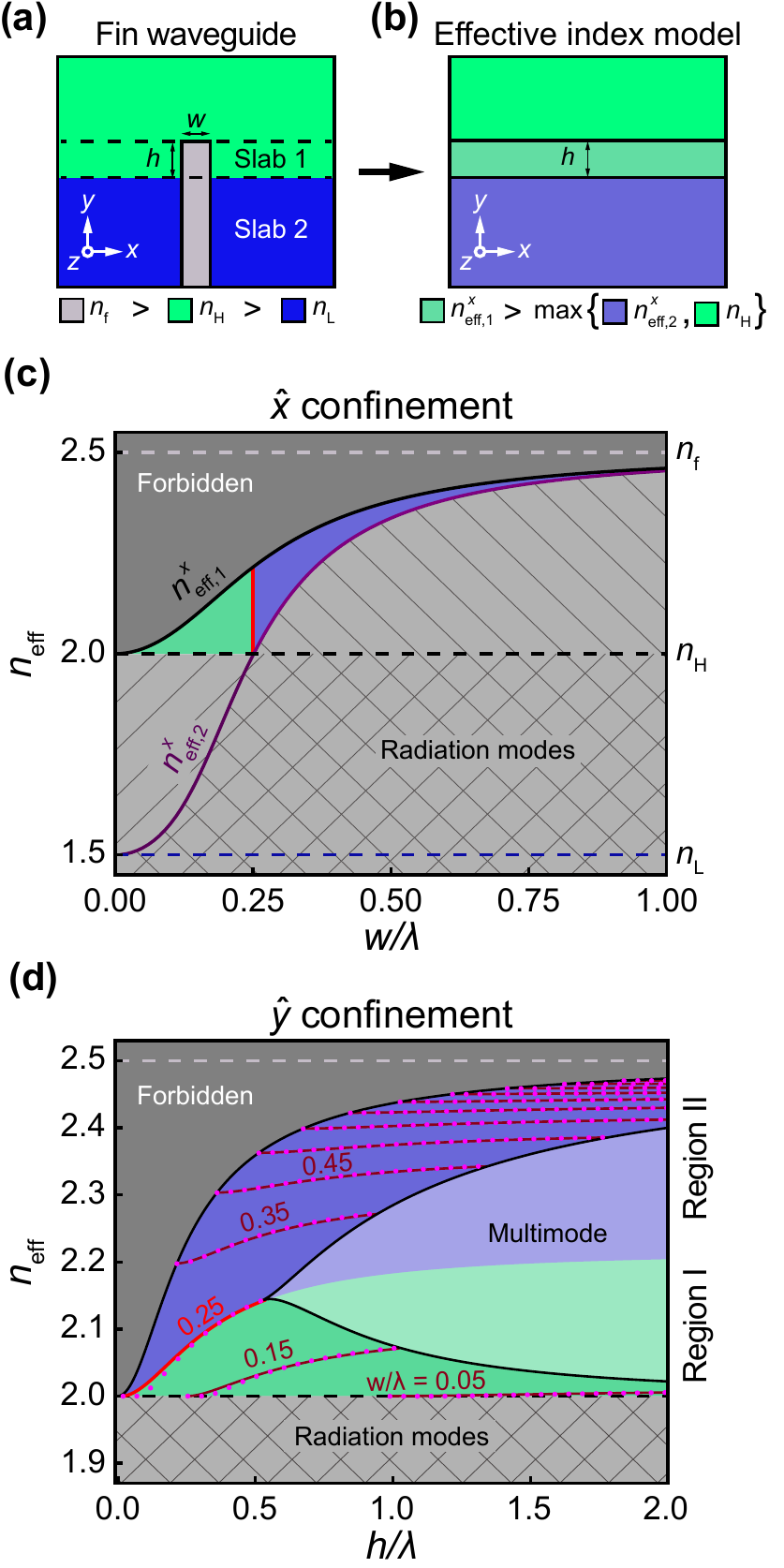}
\caption{\textbf{Generalized fin waveguide dispersion.} \textbf{(a)} Cross-section of the fin waveguide geometry.  Dashed lines mark regions that are approximated by effective indices in \textbf{(b)}.  \textbf{(c)} Dispersion of the fin waveguide as a function of $w/\lambda$.  The region in which modes are confined (blue and green shading) is bound by the effective indices of the slab waveguides in \textbf{(a)}.  \textbf{(d)} Dispersion of the fin-waveguide as a function of $h/\lambda$.  Semi-analytical curves are plotted for specific values of $w/\lambda$, and points are the result of fully-vectorial numerical simulations.}
\label{fig2}
\end{figure}

Figures~\ref{fig2}\textbf{a}--\textbf{b} illustrate an intuitive physical picture that accounts for optical confinement in the fin waveguide geometry.  By treating the two-dimensional cross-section of the $\hat{z}$-invariant waveguide dielectric topology in Fig.~\ref{fig2}\textbf{a} as two stacked slab waveguides with horizontal ($\hat{x}$) confinement, a slab waveguide with vertical ($\hat{y}$) confinement can be formed from the slab-waveguide effective indicies in $\hat{x}$.  In Fig.~\ref{fig2}\textbf{b}, the two slab waveguides comprised of $n_{\text{H}}/n_{\text{f}}/n_{\text{H}}$ (Slab~1) and $n_{\text{L}}/n_{\text{f}}/n_{\text{L}}$ (Slab~2) from Fig.~\ref{fig2}\textbf{a} are replaced by homogeneous layers in $\hat{x}$ with the effective indices of the lowest-order supported modes $n^x_{\text{eff,}1}$ and $n^x_{\text{eff,}2}$, respectively, where the effective index, $n^x_{\text{eff,}i} = \beta^x_i/k_0$, is equal to the slab waveguide propagation constant in $\hat{z}$, $\beta^x_i$, of Slab~$i$ divided by the free space wavenumber, $k_0 = 2\pi/\lambda$.  The dispersion curves of the two slabs as a function of the normalized waveguide width, $w/\lambda$, are shown in Fig.~\ref{fig2}\textbf{c} with example refractive index values of $n_{\text{f}} = 2.5$, $n_{\text{H}} = 2.0$, and $n_{\text{L}} = 1.5$. 

This treatment is key to understanding the nature of confinement in the fin structure: a fin mode exists when the effective index of the two-dimensionally confined structure, $n_{\text{eff}}$, satisfies the condition: $n^x_{\text{eff,}1} > n_{\text{eff}} > \max\{n^x_{\text{eff,}2},n_{\text{H}} \}$, as indicated by the green and blue shaded regions in Fig.~\ref{fig2}\textbf{c}.  When this condition is not met, the confined modes are degenerate with a continuum of radiation modes and become leaky, as indicated by the gray hatched region in Fig.~\ref{fig2}\textbf{c}.  As a consequence of the effective index confinement in the fin structure, only a single-mode in $\hat{x}$ is supported for the refractive index values chosen in Fig.~\ref{fig2}\textbf{c}.  Higher-order modes can be confined for a different choice of material indices, but only if higher-order modes of Slab~1 are contained in the blue or green shaded region of Fig.~\ref{fig2}\textbf{c}, \ie, they have an effective index larger than the lowest-order mode of Slab~2.

The supported modes of the two-dimensionally-confined structure are found by solving for the modes of the $\hat{y}$-confined slab waveguide in Fig.~\ref{fig2}\textbf{b}.  The resulting fin mode dispersion as a function of normalized waveguide height, $h/\lambda$, (Fig.~\ref{fig2}\textbf{d}) has two distinct regions that depend on the fin width, $w$: 

\begin{description}
\centering
 \item[\textbf{Region I}] $n^x_{\text{eff,}2} < n_{\text{H}}$; $w < w_{\text{symm}}$ (green shading), 
 \item[\textbf{Region II}] $n^x_{\text{eff,}2} > n_{\text{H}}$; $w > w_{\text{symm}}$ (blue shading). 
\end{description}

\noindent The boundary between these two regions occurs when $n^x_{\text{eff,}2} = n_{\text{H}}$ at a width that we label $w = w_{\text{symm}}$, which is indicated by a vertical red line in Fig.~\ref{fig2}\textbf{c} and the red dispersion curve ($w_{\text{symm}}/\lambda = 0.25$) in Fig.~\ref{fig2}\textbf{d}.  For an asymmetric slab waveguide, the higher-refractive-index cladding determines both the cut-off condition and the effective mode width.  In \textbf{Region~I}, the properties of the asymmetric waveguide are determined by $n_{\text{H}}$, while in \textbf{Region~II}, this role is taken by $n^x_{\text{eff,}2}$.  The change in cut-off condition between the two regions causes the inflection point in the the cut-off height (boundary for the multimode region in Fig.~\ref{fig2}\textbf{d}) at $w_{\text{symm}}$.  While the fin waveguide can only be single or few mode in width, it can be multimode in height, as indicated in Fig.~\ref{fig2}\textbf{d}.      

\begin{figure}[t!]
\centering
\includegraphics[scale=1]{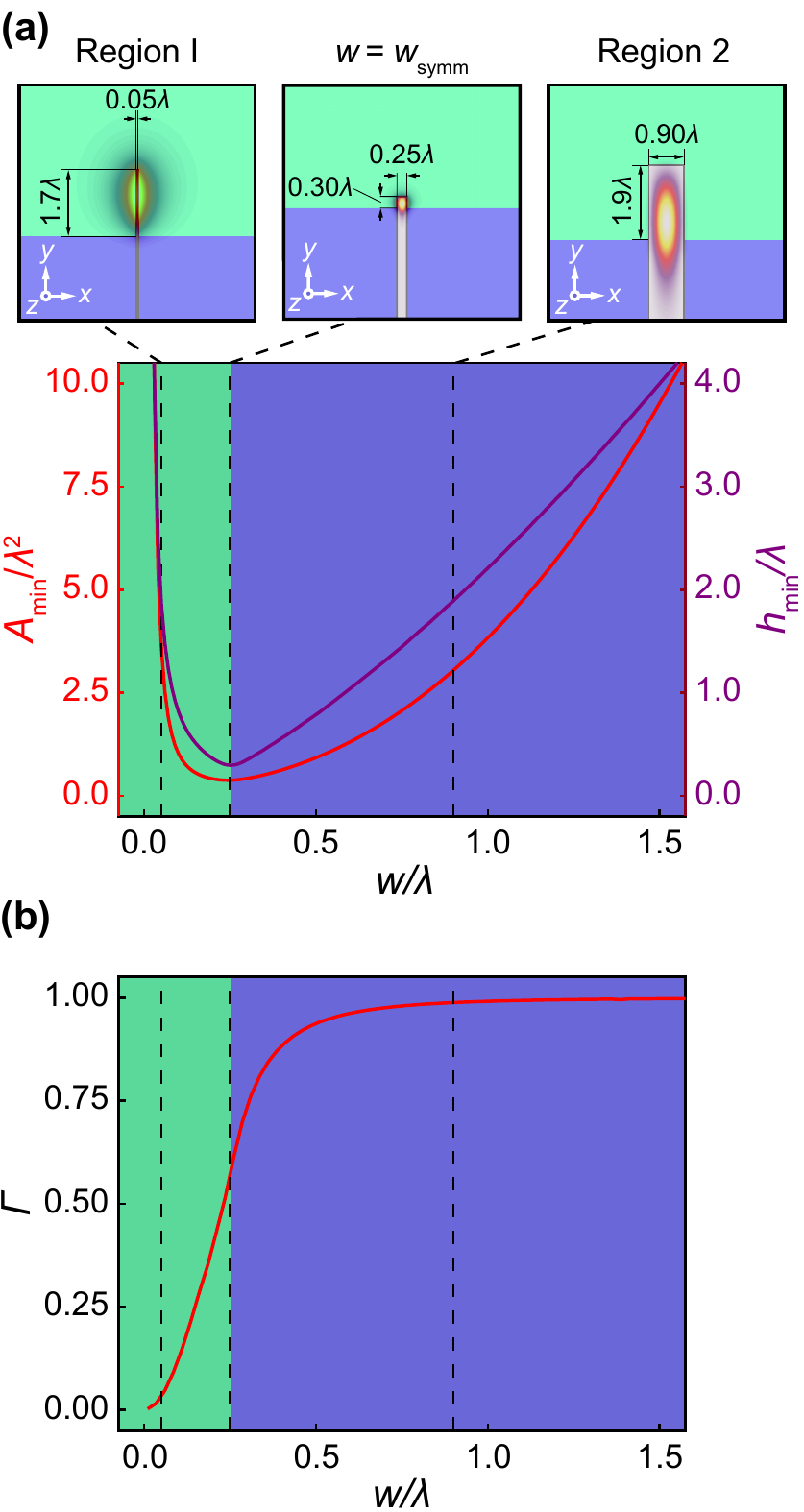}
\caption{\textbf{Optical confinement of the fin waveguide.}  \textbf{(a)} Minimum mode area as a function of $w/\lambda$ and corresponding $h_{\text{min}}/\lambda$.  Mode profiles at the three values of $w/\lambda$ marked by dashed lines are plotted above. The smallest achievable mode area occurs when $n_{\text{H}} = n^x_{\text{eff,}2}$ at $w_{\text{symm}}/\lambda = $~0.25. \textbf{(b)} The confinement factor, $\Gamma$, measures the geometric overlap between the optical mode and the high-index material.}
\label{fig3}
\end{figure}

The dispersion curves provide the allowable geometry and wavelength at which confined modes are supported.  Within these constraints, the mode area, $A_{\text{eff}}$, and confinement factor, $\Gamma$, which quantifies the overlap between the optical mode and the guiding material, provide useful design metrics for maximizing light-matter interactions in fin waveguides (see Supplementary Section 1 for definitions of these parameters).  In Fig~\ref{fig3}\textbf{a}, we calculate $A_{\text{eff}}$ for a series of $w/\lambda$, and plot the minimum, $A_{\text{min}}$, along with the corresponding height, $h_{\text{min}}/\lambda$.  Mode intensity profiles for three values of $w/\lambda$ are also shown in Fig.~\ref{fig3}\textbf{a}, and the corresponding $\Gamma$ at $A_{\text{min}}$ is plotted in Fig.~\ref{fig3}\textbf{b}.  

In \textbf{Region~I} most of the field penetration occurs in the confinement layer, relaxing the requirements on the buffer layer thickness for low leakage at the expense of reduced $\Gamma$.  Conversely, in \textbf{Region II} $A_{\text{min}}$ increases with $w/\lambda$, $\Gamma$ approaches unity, and the mode extends within the fin into the buffer layer. Waveguides designed in this region may be desirable for high-power applications.  The tightest confinement (smallest $A_{\text{min}}$) occurs at $w_{\text{symm}}$, which also corresponds to the maximum group index, $N_{\text{g}} = c/v_{\text{g}}$, where $c$ is the vacuum speed of light and $v_{\text{g}}$ is the modal group velocity (see Supplementary Section 1), making $w = w_{\text{symm}}$ an ideal design criterion for applications in non-linear or quantum optics.  The waveguide properties in Figs.~\ref{fig2} and \ref{fig3} have been calculated for the lowest-order mode with the dominant electric field component along $\hat{x}$.  Discussion of higher order modes and further details of our semi-analytical and numerical calculations are provided in Supplementary Section 1. 

\begin{figure}[t!]
\centering
\includegraphics[scale=1]{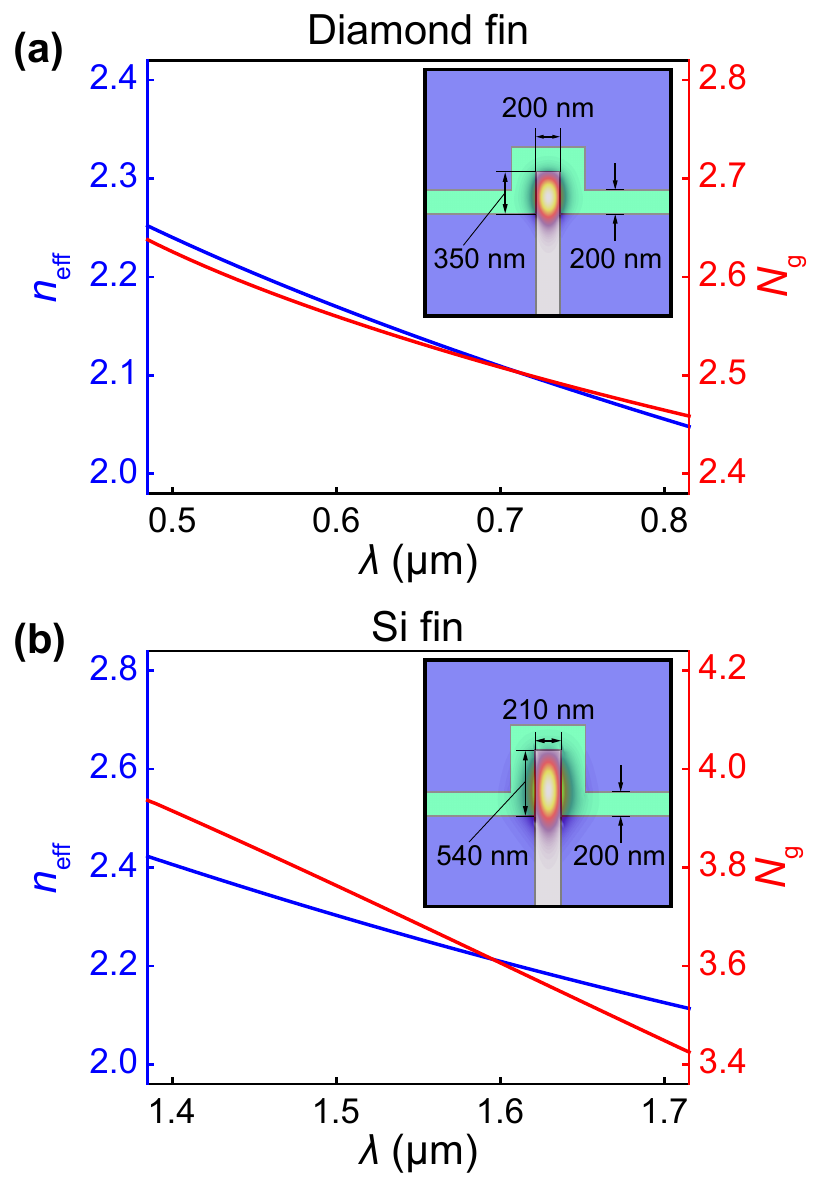}
\caption{\textbf{Design examples.}  Fin waveguides designed for maximum confinement in \textbf{(a)} diamond at $\lambda = $~\SI{637}{\nano\meter} and \textbf{(b)} silicon at $\lambda = $~\SI{1.55}{\micro\meter}.  For both structures, the effective index and group index are plotted versus $\lambda$ around the design wavelength, with the cross-sectional geometry and mode profile inset.}
\label{fig4}
\end{figure}

 To illustrate the potential of the fin waveguide, we explore geometries in two important material platforms for PICs: diamond and silicon.  The diamond waveguide is designed with an SiO$_2$ buffer layer, a conformal \SI{200}{\nano\meter}-thick Si$_3$N$_4$ confinement layer, and SiO$_2$ overcladding for single-mode operation at $\lambda = $~\SI{637}{\nano\meter}.  The operating wavelength corresponds to the nitrogen-vacancy center zero phonon line \cite{Faraon_NP_11}, which is used to achieve coherent spin-light interactions \cite{Buckley_Science_10} and distributed entanglement \cite{Bernien_Nature_13} between diamond spins.  We design the waveguide for the tightest confinement with $w = w_{\text{symm}}$ as discussed above.  The waveguide dispersion curve and group index is shown in Fig.~\ref{fig4}\textbf{a}, along with the waveguide dimensions and calculated mode intensity profile at $\lambda = $~\SI{637}{\nano\meter} in the inset. We calculate that for a buffer layer thickness exceeding \SI{1.0}{\micro\meter} the propagation loss due to substrate leakage is $<$~\SI[per-mode=symbol]{0.15}{\decibel\per\centi\meter}, which is small enough that scattering due to fabrication imperfections in a realistic device would be expected to dominate.  The bending loss for a bend radius of \SI{10}{\micro\meter} is determined to be $<$~\SI[per-mode=symbol]{0.06}{\decibel} per $90^{\circ}$ bend with a buffer layer thickness of \SI{2.5}{\micro\meter}, corresponding to an unloaded $Q$ exceeding 30,000 for a 20~\si{\micro\meter}-diameter ring resonator.  

Similarly, the silicon waveguide depicted in the inset to Fig.~\ref{fig4}\textbf{b} is designed for minimum mode area at $\lambda = $~\SI{1.55}{\micro\meter} for telecommunications applications.  With a buffer layer thickness exceeding \SI{1.5}{\micro\meter}, the propagation loss due to substrate leakage is calculated to be $<$~\SI[per-mode=symbol]{0.1}{\decibel\per\centi\meter}.  The bending loss for a bend radius of \SI{10}{\micro\meter} is determined to be $<$~\SI[per-mode=symbol]{0.1}{\decibel} per $90^{\circ}$ bend with a buffer layer of \SI{2.5}{\micro\meter}, corresponding to an unloaded $Q$ exceeding 10,000 for a ring resonator.  Additional details regarding the design and modeling of fin waveguides are provided in Supplementary Section 2.

The design strategy described in this paper provides a straightforward pathway towards the implementation of a full PIC architecture based on monolithic fin waveguides.  The required aspect ratios of the fins in Fig.~\ref{fig4} can be achieved through dry etching \cite{Wu_JAP_10, Hausmann_DRM_10}, and the dielectric stack can be fabricated using standard deposition, planarization, and lithography techniques.  Although the examples in Fig.~\ref{fig4} use a SiO$_2$/Si$_3$N$_4$ dielectric stack, the fin waveguide can be designed for any pair of materials with $n_{\text{H}} > n_{\text{L}}$.  Potential alternatives for the confinement layer include Al$_3$O$_2$ ($n \approx 1.8$), AlN ($n \approx 2.2$), SU-8 ($n \approx 1.5$), and Hydex ($n = 1.5$ to 1.9) \cite{Moss_NP_13}, and a full stack can also be designed using polymer layers \cite{Halldorsson_OpEx_10}, or chalcogenide glasses \cite{Eggleton_NP_11}.  The compatibility of this design with a number of different materials platforms allows for integration in a diverse range of fabrication processes.  For example, active devices for silicon PICs could be achieved through the realization of vertical $p$-$i$-$n$ junctions \cite{Liu_OpEx_07} aligned with the fin waveguide.  One challenge with the proposed architecture is the incorporation of devices that are typically multimode, such as Y-branches and grating couplers, since higher-order modes in $\hat{x}$ are leaky.  We envision a solution to this challenge in the form of supermode devices \cite{Yariv_07}, where multimode propagation is achieved by coupled arrays of single-mode waveguides.     

In summary, we have proposed a new waveguide design for native high-refractive-index substrates.  This method is compatible with standard fabrication processes and alleviates the need for a buried low-index layer, providing a potential route for CMOS-compatible co-integration of silicon photonics with VLSI electronics.  For InP-based PICs, the fin waveguide design provides an alternative to the InGaAs guiding layer, providing much higher confinement and a smaller mode area.  Furthermore, the geometry can be adapted for any high-index substrate material, which will lead to rapid development of PICs on emerging materials platforms for quantum information processing and sensing applications.

\section*{Acknowledgments}
We thank F. Aflatouni, J. B. Driscoll, S. A. Mann, A. L. Exarhos, and D. A. Hopper for helpful discussions and careful reading of the manuscript.   We acknowledge financial support for this research from the University of Pennsylvania. 

\bibliographystyle{osajnl}
\bibliography{Grote_Optica_15}

\end{document}


\title{Single-Mode Optical Waveguides on Native High-Refractive-Index Substrates: Supplementary Information}

\author{Richard R. Grote, Lee C. Bassett$^*$}

\affiliation{Quantum Engineering Laboratory, Department of Electrical and Systems Engineering, University of Pennsylvania, 200 S. 33rd Street, Philadelphia, PA 19104, USA}

\affiliation{$^*$Corresponding author: lbassett@seas.upenn.edu}

\date{\today}

\maketitle

\section{Waveguide dispersion derivation and definitions}
\label{dispersion}

The supported modes of the $\hat{z}$-invariant refractive index profile, $n(x,y)$, in Fig.~2\textbf{a} of the main text can be solved by casting Maxwell's equations as an eigenvalue problem:

\begin{equation}
\hat{A}\ket{m} = \beta_m^2 \ket{m}
\label{eig_2D}
\end{equation}

\noindent where $\beta_m = k_0 n^m_{\text{eff}}$ is the propagation constant of mode $m$, $k_0 = \frac{2\pi}{\lambda}$ is the free space wavenumber, $n^m_{\text{eff}}$ is the effective refractive index of mode $m$ and the operator $\hat{A}$ takes the following form when projected onto the position basis:

\begin{equation}
\hat{A} = \nabla^2_{\text{t}} + k_0^2 n^2(x,y).
\end{equation}

\noindent The eigenvectors represent the transverse electric fields:

\begin{equation}
\hat{A} \ket{\mathbf{E}^m_{\text{t}}} = \beta_m^2 \ket{\mathbf{E}^m_{\text{t}}}
\label{eig_2D_pos}
\end{equation}

\noindent where 

\begin{equation}
 	\ket{\mathbf{E}^m_{\text{t}}} = \begin{bmatrix} 
 										E^m_x \\ 
 										E^m_y 
 								\end{bmatrix}.
\end{equation}  

\noindent With two-dimensional confinement the two transverse field components are mixed by boundary conditions for the tangential electric fields and perpendicular displacement fields at the dielectric interfaces contained in $n^2(x,y)$.   The modes of the two-dimensionally confined structure cannot be calculated analytically; however, solutions to equations (\ref{eig_2D_pos}) can be found using numerical approaches such as finite-difference method (FDM) \cite{Fallahkhair_JLT_08}.  An alternate approach is to use the effective index method \cite{Nishihara_89, Yariv_07} to find an approximate solution by treating equation (\ref{eig_2D}) as two separable problems in $\hat{x}$ and $\hat{y}$.  We use the effective index method outlined in the following section to find approximate solutions to the fin waveguide dispersion, and verify our calculations with FDM \cite{Fallahkhair_JLT_08}.

\subsection*{Effective index method}

In the one-dimensional analysis for $\hat{x}$, $n^2(x,y) \rightarrow n_i^2(x)$, $\hat{A} \rightarrow \hat{A}_x = \frac{\partial^2}{\partial x^2} + k_0^2 n_i^2(x)$, and equation (\ref{eig_2D_pos}) reduces to two separable equations representing two orthogonal polarizations.  We represent these polarizations by $r = \hat{x}$ for the horizontal ($E_x$)-polarization, and $r = \hat{y}$ for the vertical ($E_y$)-polarization.  We follow the approach of \cite{Nishihara_89} to solve for the slab waveguide effective indices.  The indices of the $\hat{x}$ confined slab waveguides shown in Fig.~2\textbf{a} of the main text are: 

\begin{eqnarray}
\text{Slab 1: } &	n_1(x) = \begin{cases}
					n_{\text{c},1} = n_{\text{H}}, & x > \frac{w}{2} \\
					n_{\text{f}}, & \frac{w}{2} > x > -\frac{w}{2} \\
					n_{\text{c},1} = n_{\text{H}}, & -\frac{w}{2} > x
	\end{cases} \\
\text{Slab 2: } &	n_2(x) = \begin{cases}
					n_{\text{c},2} = n_{\text{L}}, & x > \frac{w}{2} \\
					n_{\text{f}}, & \frac{w}{2} > x > -\frac{w}{2} \\
					n_{\text{c},2} = n_{\text{L}}, & -\frac{w}{2} > x
	\end{cases}
\end{eqnarray}

\noindent The field is assumed to be sinusoidal in the guiding region, and exponentially decaying outside. The phase constants for the field in each region are defined as:

\begin{align}
\gamma^x_{\text{c},i} & = k_0 \sqrt{(n^x_{\text{eff},i})^2 - n_{\text{c},i}^2} \label{gammax} & \text{cladding}\\
k^x_{\text{f},i} & = k_0 \sqrt{n_{\text{f}}^2 - (n^x_{\text{eff},i})^2} \label{kx} & \text{guiding region}
\end{align}

\noindent where $i = 1,2$, corresponding to Slab~1 and Slab~2, respectively.  The effective index of Slab~$i$, $n^x_{\text{eff},i}$, is found by matching the phase constants at the boundaries to find the following eigenvalue equations:

\begin{align}
k_{x,i} w &= (p_i + 1)\pi - 2\tan^{-1}\left(\frac{k_{x,i}}{\gamma^x_{\text{c},i}} \right),  & r = \hat{y} \label{eigxTE}\\
k_{x,i} w &= (p_i + 1)\pi - 2\tan^{-1}\left(\frac{n_{\text{c},i}}{n_{\text{f}}}\right)^2 \left( \frac{k_{x,i}}{\gamma^x_{\text{c},i}} \right), & r = \hat{x} \label{eigxTM}
\end{align}

\noindent where $p_i = 0,1,2,...$ is the mode index for Slab $i$.  The phase constants are related to $\beta$ by equations (\ref{gammax}) and (\ref{kx}), and solutions to the transcendental equations (\ref{eigxTE}) or (\ref{eigxTM}) provide the propagation constant for mode $p_i$.  The cut-off condition for mode $p_i$ can be expressed in terms of a minimum waveguide width:

\begin{align}
	w_{\text{cut-off},p_i} & = \frac{\lambda \pi p_i}{2 \sqrt{n_{\text{f}}^2 - n_{\text{H}}^2}}, & p_i = 0,1,2, ...
	\label{wcutoff}
\end{align}

\noindent Since the slab waveguides in $\hat{x}$ are symmetric, there is no cut-off for the lowest order mode.  Using the solutions for the slab waveguides in Fig.~2\textbf{a} of the main text, a $\hat{y}$-confined asymmetric slab waveguide (Fig.~2\textbf{b} of the main text) can be constructed with the following indices:

\begin{eqnarray}
	n(y) = \begin{cases}
					n_{\text{H}}, & y > \frac{h}{2} \\
					n^{x}_{\text{eff},1}, & \frac{h}{2} > y > -\frac{h}{2} \\
					n^{x}_{\text{eff},2}, & -\frac{h}{2} > y
	\end{cases}
\end{eqnarray}

 \begin{figure}[t!]
\centering
\includegraphics[scale=1]{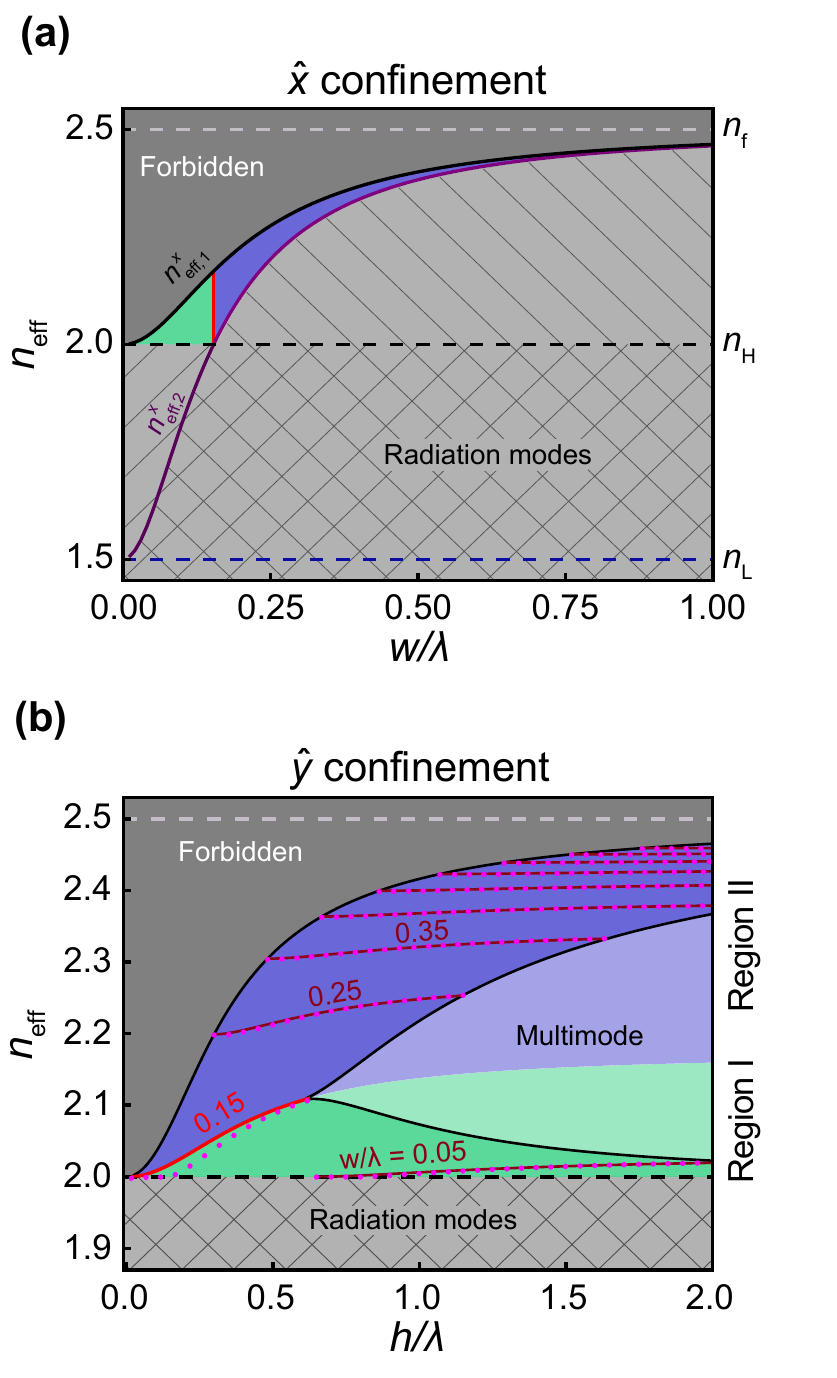}
\caption{Dispersion plots for the lowest-order $\ket{0,0,\hat{y}}$ mode of the fin waveguide in Fig.~2\textbf{a} of the main text.  }
\label{fig1}
\end{figure}

Using the nomenclature from \cite{Nishihara_89}, the asymmetric waveguide can be parameterized by a ``substrate'' with refractive index $n_{\text{s}}$ and ``cladding'' with refractive index $n_{\text{c}}$, where $n_{\text{f}} > n_{\text{s}} > n_{\text{c}}$.  The effective index of Slab~2, $n^x_{\text{eff},2}$, depends on $w/\lambda$, thus we define the ``substrate'' and ``cladding'' indices in the following way: 

\begin{align}
	n^y_{\text{s}} & = \max \left\{ n^x_{\text{eff},2}, n_{\text{H}} \right\} \\
	n^y_{\text{c}} & = \min \left\{ n^x_{\text{eff},2}, n_{\text{H}} \right\}
\end{align}

Using these definitions, the phase constants of the asymmetric waveguide are defined as follows: 

\begin{align}
\gamma^y_{\text{c}} & = k_0 \sqrt{(n^y_{\text{eff}})^2 - (n^y_{\text{c}})^2} \\
\gamma^y_{\text{s}} & = k_0 \sqrt{(n^y_{\text{eff}})^2 - (n^y_{\text{s}})^2} \\
k_y & = k_0 \sqrt{(n^x_{\text{eff},1})^2 - (n^y_{\text{eff}})^2}
\end{align}

\noindent which result in the following eigenvalue equations:

\begin{align}
	k^y_{\text{f}} h = (q + 1)\pi & - \tan^{-1}\left(\frac{k^y_{\text{f}}}{\gamma^y_{\text{s}}} \right) & \nonumber \\
	            & - \tan^{-1}\left(\frac{k^y_{\text{f}}}{\gamma^y_{\text{c}}} \right), & r = \hat{x} \\ 
	k^y_{\text{f}} h = (q + 1)\pi & - \tan^{-1}\left(\frac{n^y_{\text{s}}}{n^x_{\text{eff},1}}\right)^2 \left( \frac{k^y_{\text{f}}}{\gamma^y_{\text{s}}} \right) & \nonumber \\
			 & - \tan^{-1}\left(\frac{n^y_{\text{c}}}{n^x_{\text{eff},1}}\right)^2 \left( \frac{k^y_{\text{f}}}{\gamma^y_{\text{c}}} \right), & r = \hat{y}
\end{align}

\noindent where $q = 0,1,2,...$ is the mode index for $\hat{y}$-confinement.  The limits on allowable height for single mode operation are found by the cut-off condition for the lowest and first order modes of the asymmetric waveguide in Fig.~\ref{fig2}\textbf{d}, which define the asymmetry parameter $a^y$:

\begin{equation}
	a^y = \begin{cases}
			\frac{(n^y_{\text{s}})^2 - (n^y_{\text{c}})^2}{(n^x_{\text{eff},1})^2 - (n^y_{\text{s}})^2}, & r = \hat{x} \\
			 \left( \frac{n^x_{\text{eff},1}}{n^y_{\text{c}}} \right )^4 \frac{(n^y_{\text{s}})^2 - (n^y_{\text{c}})^2}{(n^x_{\text{eff},1})^2 - (n^y_{\text{s}})^2}, & r = \hat{y}
			\end{cases}
			\label{asymmetry_param}
\end{equation}

 \noindent The cut-off for mode $\ket{0,q,r}$ occurs at a height of:
 
 \begin{align}
 	h_{\text{cut-off},q} & = \frac{ \lambda \left(\tan^{-1} \sqrt{a^y} + q\pi \right) }{2 \pi \sqrt{(n^x_{\text{eff},1})^2 - (n^y_{\text{s}})^2}}, & q = 0,1,2,... \label{hcutoff}
 \end{align}

Although pure polarization state solutions of the two-dimensionally confined modes do not exist, we use the mode numbers of the approximate effective index method solutions as waveguide eigenvalue labels:

\begin{equation}
\ket{m} \approx \ket{p,q,r}.
\end{equation}

All of the calculations in the main text have been performed for the $\ket{0,0,\hat{x}}$ mode.  The dispersion curves for the $\ket{0,0,\hat{y}}$ mode of the structure in Fig.~2\textbf{a} of the main text are shown in Fig.~\ref{fig1}.

In the example considered in Fig.~2 of the main text, the fin waveguide only supports a single horizontal mode ($p=0$).  For a different choice of $n_{\text{L}}$ and $n_{\text{H}}$, such that the cut-off width of higher-order modes in the guiding layer is smaller than $w_{symm}$, additional horizontal modes can be supported, but the $p = 0$ mode for the buffer region always provides a lower limit for $n_{\text{eff}}$ of confined modes as shown in Fig.~2\textbf{c} of the main text.  

\subsection*{Power flow}

The complex Poynting vector is defined as \cite{Saleh_07}:

\begin{equation}
S = - \frac{1}{2} \mathbf{E} \times \mathbf{H}^*.
\label{poynting_vec}
\end{equation}

\noindent The $\hat{z}$-component of the Poynting vector represents the mode intensity profiles plotted in the main text and is defined as: $\text{Re}\left\{S_z \right\} = \text{Re}\left\{S\cdot \hat{z} \right\}$.  The time-averaged power flow in $\hat{z}$ can also be found from the complex Poynting vector:

\begin{equation}
P_z = \int \text{Re}\left\{S_z\right\} \mathrm{d}A
\label{P_z}
\end{equation}

\noindent where $A$ is area.  For all calculations the modes are normalized such that $P_z = $~\SI{1}{\watt}.  

For substrate loss calculations, there is power flow in $\hat{y}$, which is visualized by the $\hat{y}$-component of the Poynting vector, $\text{Re}\left\{S_y \right\} = \text{Re}\left\{S\cdot \hat{y} \right\}$, plotted in the insets of Fig.~\ref{fig3}.  The time-averaged power flow in $\hat{y}$ is defined in a similar manner to equation (\ref{P_z}).

\subsection*{Mode area and confinement factor}

For small perturbations the effect of waveguiding on light-matter interaction can be approximated as \cite{Joannopoulos_11, Grote_14}:

\begin{equation}
\Delta n_{\text{eff}} \approx -\frac{\Delta n}{n} N_{\text{g}} F
\end{equation}

\noindent where $\Delta n_{\text{eff}}$ is the change in the waveguide effective index due to a perturbation $\Delta n$, which depends on both the group index, $N_{\text{g}}$, and the fraction of mode energy contained in the perturbed region,  $F$.  The perturbation can be complex valued and can represent light-matter interactions such as absorption, gain, or material nonlinearity.  Since we are concerned with perturbations to the waveguide core, $F$ can be approximated by the confinement factor, $\Gamma$, which is defined as:

\begin{equation}
\Gamma = \frac{\int_{n_{\text{f}}} \text{Re} \left\{S_z\right\} \mathrm{d}A}{\int \text{Re} \left\{S_z\right\} \mathrm{d}A}
\end{equation}

\noindent and is plotted in Fig.~3\textbf{b} of the main text.

The ratio of the group velocity, $v_{\text{g}}$, to phase velocity, $v_{\text{p}}$, can be related to electric and magnetic field energies in the following way \cite{Loh_PRE_09}:

\begin{figure}[t!]
\centering
\includegraphics[scale=1]{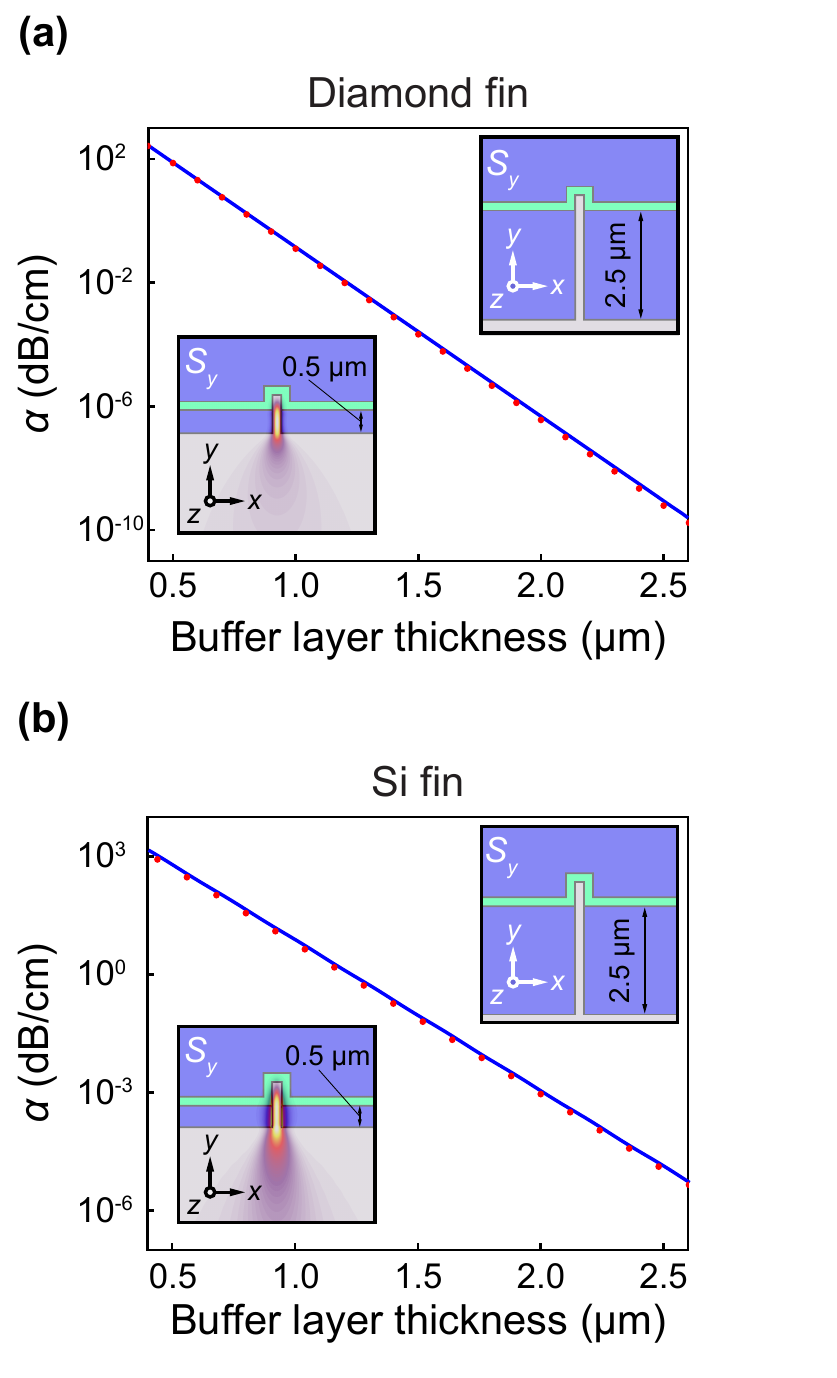}
\caption{ \textbf{(a)},\textbf{(b)} Propagation loss, $\alpha$, versus buffer layer thickness at the design wavelength.  Solid lines and points represent two different calculation methods: absorbing boundary conditions (blue line) and coupled-mode theory (red points). \textbf{(inset)} Poynting vector in the $y$-direction, $\text{Re}\left\{S_y \right\}$, for buffer layer thicknesses of \SI{0.5}{\micro\meter} and \SI{2.5}{\micro\meter}.}
\label{fig3}
\end{figure}

\begin{equation}
\frac{v_{\text{g}}}{v_{\text{p}}} = F_{\text{t}} - F_z
\label{vg_energy}
\end{equation}

\noindent where $F_{\text{t}}$ is the fraction of mode energy contained in the transverse fields, $E_x, E_y, H_x, H_y$, and $F_z$ is the fraction of mode energy contained in the longitudinal fields, $E_z, H_z$.  As was noted in \cite{Driscoll_OpEx_09}, $N_{\text{g}}$ is inversely proportional to the effective mode area, $A_{\text{eff}}$, as related through the energy contained in the longitudinal fields.  Thus, $A_{\text{eff}}$, can be used as a proxy for $N_{\text{g}}$, and the minimum mode area, $A_{\text{min}}$, corresponds to the maximum $N_{\text{g}}$.  

The effective mode area, $A_{\text{eff}}$, plotted in Fig.~3\textbf{a} of the main text has been calculated as the area of an ellipse with axes defined by the modal effective width, $w_{\text{eff}}$ and effective height, $h_{\text{eff}}$, as follows:
 
\begin{equation}
A_{\text{eff}} = \frac{\pi}{4} w_{\text{eff}} h_{\text{eff}}
\end{equation}

\noindent where $w_{\text{eff}}$ and $h_{\text{eff}}$ are calculated using the definitions in Supplementary Section \ref{dispersion}:
 
\begin{align}
w_{\text{eff}} & = w + \frac{2}{\gamma^x_{\text{c},1}} \\
h_{\text{eff}} & = h  + \frac{1}{\gamma^y_{\text{s}}} + \frac{1}{\gamma^y_{\text{c}}}
\end{align}

\begin{table}[t!]
\centering
\caption{\bf Sellmeier equation parameters }
\begin{tabular}{ccccc}
\hline
Material & $i$ & $A_i$ & $\lambda_i$~[\si{\micro\meter}] & Ref. \\
\hline
Diamond	& 1	& 0.3306	& 0.175	& \cite{Mildren_13} \\
			& 2	& 4.3356	& 0.106	&	\\
\hline
			& 1	& 10.6684		& 0.3015		&	\\
Si			& 2	& 0.0030		& 1.1347		&   \cite{Saleh_07} \\ 
			& 3 & 1.5413		& 1104		&   \\
\hline
Si$_3$N$_4$	& 1	& 2.8939	& 0.13967	&  \cite{Baak_AO_82} \\
\hline
			& 1	& 0.6962	& 0.06840	& \\
SiO$_2$	& 2 & 0.4079	& 0.1162	& \cite{Saleh_07} \\
			& 3	& 0.8975	& 9.8962	& \\
\hline
\end{tabular}
  \label{Sellmeier}
\end{table}

\begin{figure}[t!]
\centering
\includegraphics[scale=1]{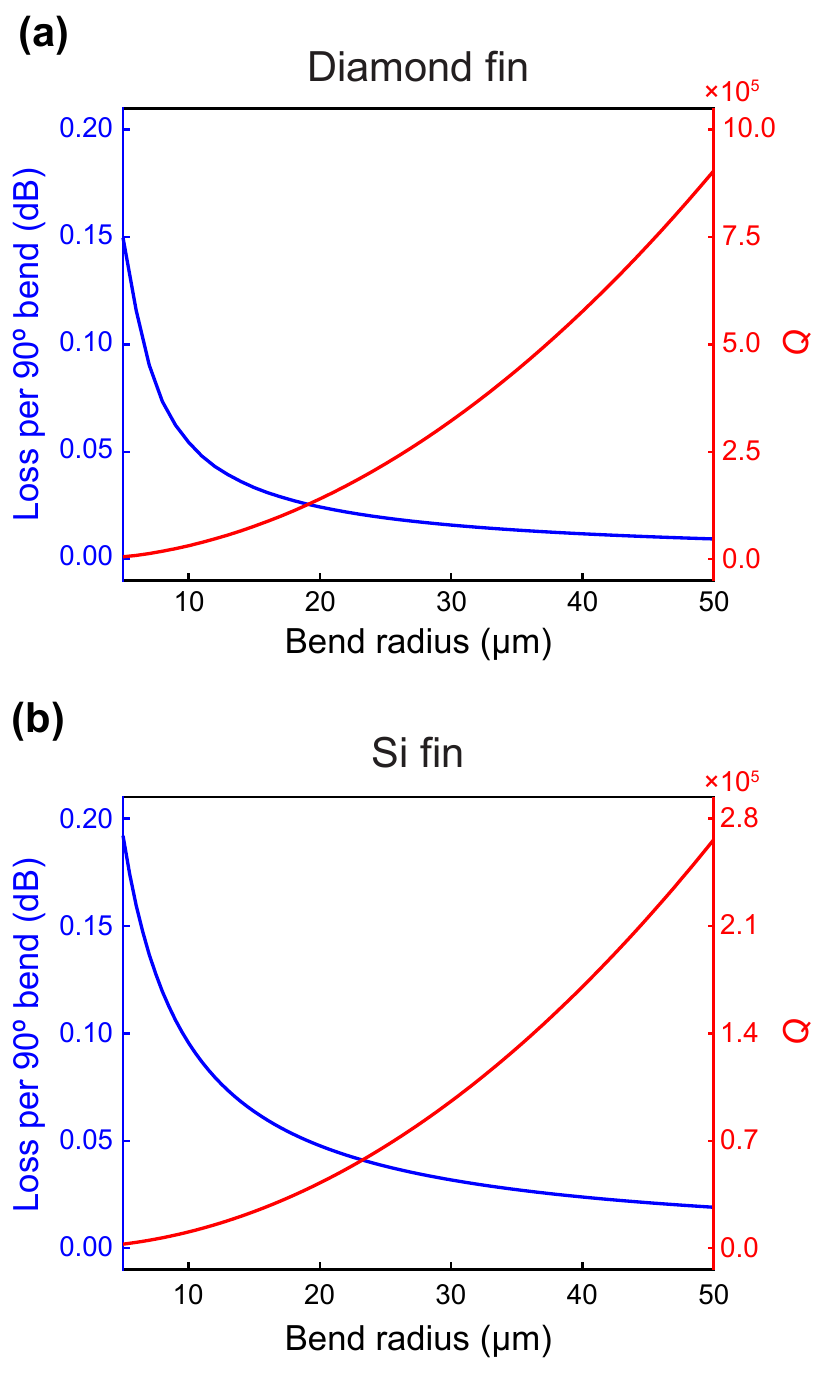}
\caption{Bending loss and $Q$-factor as a function of bend radius for \textbf{(a)} diamond fin and \textbf{(b)} silicon fin.}
\label{fig2}
\end{figure}

\section{Diamond and silicon fin waveguide designs}

\subsection*{Substrate leakage}

The substrate leakage as a function of buffer layer thickness shown in Fig.~\ref{fig3} has been calculated using two different methods: coupled-mode theory (red points) and absorbing boundary conditions (blue line).  For the coupled-mode theory calculations the unperturbed mode is calculated using FDM without the high index substrate.  Coupling of the unperturbed mode to the substrate is then calculated by using the following overlap integral \cite{Nishihara_89,Yariv_07}:

\begin{equation}
\alpha_{\text{eff}} = \braket{\mathbf{E}|\Delta \varepsilon | \mathbf{E}} = \frac{\omega \varepsilon_0}{4} \int \mathbf{E}^* \cdot \Delta n^2(x,y) \mathbf{E} ~ \mathrm{d}A  
\end{equation}

\noindent where $\Delta n^2(x,y) = n_{\text{f}}^2$ and the overlap region extends from the bottom of the buffer layer to $-\infty$ in $\hat{y}$, and from $-\infty$ to $+\infty$ in $\hat{x}$. 

For the second method, the high-index substrate is included in the FDM calculation, and perfectly matched layer (PML) boundary conditions are added to the simulation cell. These boundary conditions allow for absorption at the simulation cell edge with minimal reflections. The FDM solver can find complex eigenvalues \cite{Fallahkhair_JLT_08}, which can be related to the propagation loss in the following manner:

\begin{equation}
\alpha_{\text{eff}} = \frac{4 \pi \mathrm{Im} \{n_{\text{eff}} \}}{\lambda}
\end{equation}

\subsection*{Bending loss}
Bending loss of the structures in Fig.~\ref{fig2} is calculated in cylindrical coordinates with FDM using the method in \cite{Lui_JLT_98}, where the loss per 90$^{\circ}$ bend as a function of bend radius is shown.  A buffer layer thickness of $t_B = $~\SI{2.5}{\micro\meter} has been used for both waveguides.  The bending-loss limited $Q$-factor of a ring resonator is also shown in Fig.~\ref{fig2}, where $Q = \omega/\gamma$, $\omega = ck_0$ is the center frequency and $\gamma = \frac{\alpha_{\text{eff}} c}{N_{\text{g}}}$.  The values in Fig.~\ref{fig2}\textbf{a,b} are calculated at $\lambda = $~\SI{637}{\nano\meter} and $\lambda = $~\SI{1.55}{\micro\meter}, respectively.

\subsection*{Material Sellmeier equations}

The calculations in the main text use Sellmeier equations to model the various material refractive indices.  The parameters for each material are given in table \ref{Sellmeier} and are used in the following equation \cite{Saleh_07}:

\begin{equation}
n^2 = 1 + \sum_i \frac{A_i \lambda^2}{\lambda^2 - \lambda_i^2}
\end{equation} 

For the designs in Fig.~4 of the main text, the diamond material stack has the following refractive indices at \SI{637}{\nano\meter}: $n_{\textrm{diamond}} = 2.41$, $n_{\textrm{Si}_3\textrm{N}_4} = 2.01$, and $n_{\textrm{SiO}_2} = 1.46$ as determined from Sellmeier equations for each material.  The silicon material stack has the following refractive indices at $\lambda = $~\SI{1.55}{\micro\meter}: $n_{\textrm{Si}} = 3.48$, $n_{\textrm{Si}_3\textrm{N}_4} = 1.98$, and $n_{\textrm{SiO}_2} = 1.44$.
 
\bibliographystyle{osajnl}
\bibliography{Grote_Optica_15}